\documentclass[preprintnumbers,nofootinbib,showpacs,eqsecnum,pre,12pt]{revtex4-1}
\usepackage{amsmath}         
\usepackage{amssymb}          
\usepackage[pdftex]{graphicx}	
\usepackage[pdftex]{hyperref} 
\usepackage[utf8x]{inputenc}
\usepackage{color}
\usepackage{ulem}
\usepackage{afterpage}
\usepackage{changepage}
\usepackage{setspace}
\usepackage{ragged2e}

\def\gsim{\raise0.3ex\hbox{$\;>$\kern-0.75em\raise-1.1ex\hbox{$\sim\;$}}}
\def\lsim{\raise0.3ex\hbox{$\;<$\kern-0.75em\raise-1.1ex\hbox{$\sim\;$}}}

\begin{document}
\begin{center}

\large{\textbf{Resolving a challenging supersymmetric low-scale seesaw scenario at the ILC}}

\small {J. Masias}

\textit{
Secci\'on F\'isica, Departamento de Ciencias,}

\textit{
Pontificia Universidad Cat\'olica del Per\'u, Apartado 1761, Lima, Peru
}

\small {Email: j.masias@pucp.edu.pe}

\end{center}
\begin{adjustwidth}{35pt}{35pt}

\singlespacing{
\small{
We investigate a scenario inspired by natural supersymmetry, where neutrino data is explained within a low-scale seesaw scenario. For this the Minimal Supersymmetric Standard Model is extended by adding light right-handed neutrinos and their superpartners, the R-sneutrinos. Moreover, we consider the lightest electroweakinos to be higgsino-like.

We demonstrate that a prospective international $e^+ e^-$ linear collider with a center of mass energy of 1 TeV will be able to discover sleptons in scenarios which can be difficult for current colliders. Moreover, we also show that a measurement of the spectrum will be possible within 1-3 per-cent accuracy.}}\\
\end{adjustwidth}
\begin{center}
    \textit{Talk presented at the International Workshop on Future Linear Colliders (LCWS2021), 15-18 March 2021. C21-03-15.1.}
\end{center}
\section{Introduction}

\linespread{1.3}
\small{I}n a previous work we investigated a supersymmetric low scale seesaw model at the LHC~\cite{Cerna-Velazco:2017cmn}. Here we considered the SUSY partners of right handed neutrinos, the R-sneutrinos, to be the lightest supersymmetric particles (LSPs). These are followed in mass by the L-sneutrinos and the charged sleptons. This mass hierarchy leads to slepton decays containing SM bosons instead of the naively expected leptons.

For this kind of models, we investigate the possible reach of the International Linear Collider (ILC), running at $\sqrt{s}=1$ TeV. In particular, we find the required luminosities in order to have a discovery. Finally, we show that precise measurements of the slepton masses are possible at the ILC (see \cite{Masias:2021uga} for a further discussion).
\section{Scenarios of Interest}
\small{We} start with the MSSM superpotential, extended by three singlet superfields $\hat\nu_{R}$, such that R-parity is conserved:
\begin{align}
\mathcal{W}_{{\rm eff}} = \mathcal{W}_{\rm MSSM}
+ \frac{1}{2} (M_R)_{ij}\,\hat{\nu}_{Ri}\,\hat{\nu}_{Rj}
+ (Y_\nu)_{ij}\,\widehat{L}_i \cdot \widehat{H}_u\, \hat{\nu}_{Rj}
\end{align}
With the following allowed soft SUSY-breaking terms:
\begin{equation}
\mathcal{V}^{soft} =\mathcal{V}_{\rm MSSM}^{soft}
  + (m^2_{\tilde\nu_R})_{ij}\tilde{\nu}^*_{Ri}\tilde{\nu}_{Rj}
  + \bigg(\frac{1}{2}(B_{\tilde\nu})_{ij}\tilde{\nu}_{Ri}\tilde{\nu}_{Rj} + (T_\nu)_{ij}\,\tilde{L}_i \cdot H_u\, \tilde{\nu}_{Rj} + \text{h.c.} \bigg)
\end{equation}

Based on naturalness arguments~\cite{Papucci:2011wy,Hall:2011aa}, we assume $\mu\ll M_{1,2,3}$ such that the lightest neutralinos and chargino are higgsino-like. For simplicity we take $B_{\tilde\nu}=T_{\nu}=0$, assume diagonal slepton soft masses and fix $\mu=500$ GeV, $\tan\beta=6$. We also assume $m_{\tilde{\nu}_R}\in[0,200]$ GeV and $m_{\tilde{L}}=m_{\tilde{E}}\in[100,450]$ GeV, such that $m_{\tilde{\nu}_R}<m_{\tilde{L}}< \mu $. All other SUSY content effectively decoupled.

Given this spectrum, the first two generations of sleptons will have an off-shell (on-shell) decay $\tilde{\ell}\xrightarrow{}W^*\tilde{\nu}_L$ ($\tilde{\ell}\xrightarrow{}W \tilde{\nu}_R$) with $\sim 90\%$($\sim 10\%)$ probability, whereas the $\tilde{\tau}_1$($\tilde{\nu}_L$) will have a $\sim 100\%$ on-shell decay $\tilde{\tau}_1\xrightarrow{}W \tilde{\nu}_R$ ($\tilde{\nu}_L\xrightarrow{}Z/H \tilde{\nu}_R$). These decay channels become dominant as soon as the phase-space allows it. We are considering three main scenarios:
\begin{itemize}
    \item Scenario SE, the only light MSSM sleptons are the $\tilde e_L,\,\tilde e_R$ and $\tilde\nu_{eL}$~\cite{Jones-Perez:2013uma}.
    
    \item Scenario ST, $\tilde\tau_1,\,\tilde\tau_2$ and $\tilde\nu_{\tau L}$ are the light MSSM sleptons, common in split-family SUSY~\cite{Cohen:1996vb,Craig:2011yk,Delgado:2011kr,Larsen:2012rq,Craig:2012hc,Blankenburg:2012nx}. Here the decays are similar to those of light electroweakinos ~\cite{Battaglia:2006bv}, so we expect electroweakino searches to be sensitive to this scenario. 
    
    \item Scenario DEG, all MSSM sleptons have the same soft mass, $m_{\tilde L}=m_{\tilde e_R}$.
    
\end{itemize}
\section{Results at ILC}
We consider a 1 TeV ILC run with Type \textbf{B} polarization ($e^-_Re^+_L$) and we set the electrons (positrons) to be 80\% (20\%) polarized \cite{Baer:2013cma}. Our channels of interest are stau ($\tilde{\tau}_1^-\tilde{\tau}_1^+$) and sneutrino ($\tilde{\nu}_L\tilde{\nu}_L$) production. As final states we will have two, possibly different, SM bosons and missing energy. We consider the following cuts \cite{Masias:2021uga}, which are modified from \cite{Suehara:2009bj}:
\begin{itemize}
    \item Missing transverse momentum $p_{T}^{miss}> 50$ GeV
\item Exactly four jets or b-jets with $p_T> 20$ GeV,  each 
    \item Two reconstructed SM bosons such that their invariant masses satisfy:
    \begin{equation}
\dfrac{(m_1-{m_B}_1)^2+(m_2-{m_B}_2)^2}{\sigma^2}<4,\hspace{1cm} \sigma = 5 \text{ GeV}
\end{equation}
    \item No leptons with $p_T> 25$ GeV.
\item The angle between the missing momentum and the beam line must satisfy  $|\cos(\theta_{\rm miss})|<0.99$
\end{itemize}

We then apply these cuts to our signal and background. In Figure 1 we present the required luminosity in order to obtain a 5$\sigma$ discovery for our three scenarios.\\

\begin{figure}[h]
\centering
\includegraphics[width=0.99\textwidth]{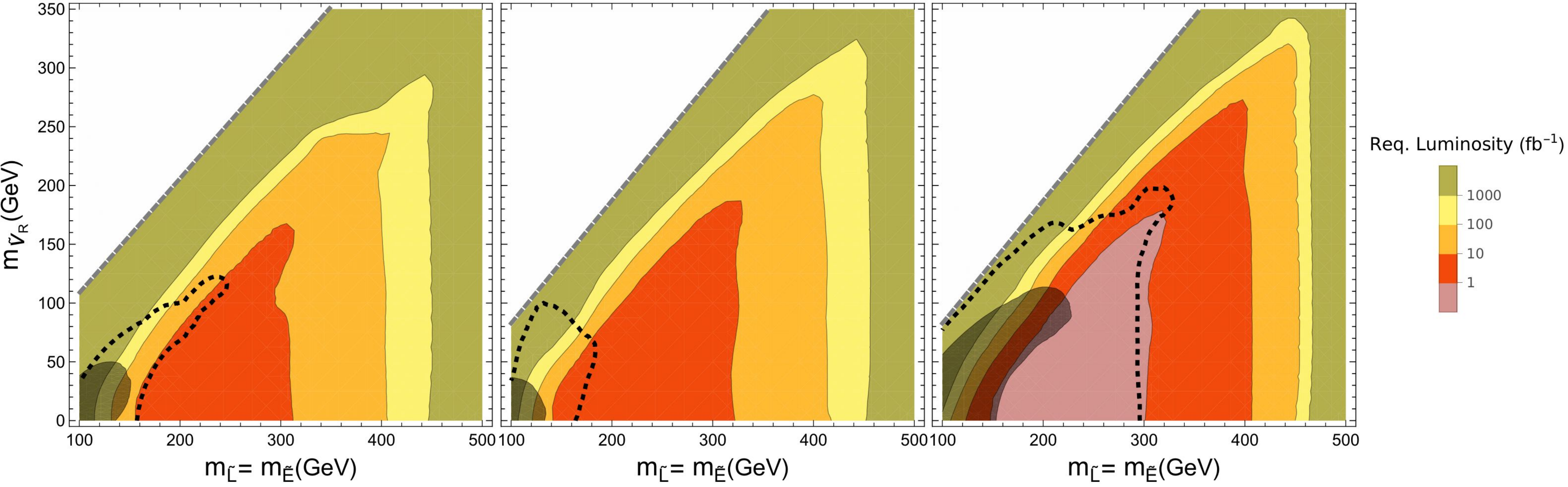}
\caption{Necessary luminosity, in fb$^{-1}$, to achieve $s/\sqrt{b}=5$. Scenarios SE, ST and DEG are shown left to right. Luminosities lower than 1, 10, 100 and 1000~fb$^{-1}$ are shown in pink, dark orange, light orange and yellow, respectively. The shaded region indicates the current LHC exclusion. Similarly, the dashed lines indicate the expected LHC reach for 300~fb$^{-1}$.}
\label{fig:all_lum}
\end{figure}

As we can see for all our scenarios, the ILC expectation covers a large part of the parameter space previously unprobed by the LHC. We can probe slepton masses up to $450$ GeV with a very conservative luminosity.\\

\begin{figure}[t]
	\centering
\includegraphics[width=0.99\textwidth]{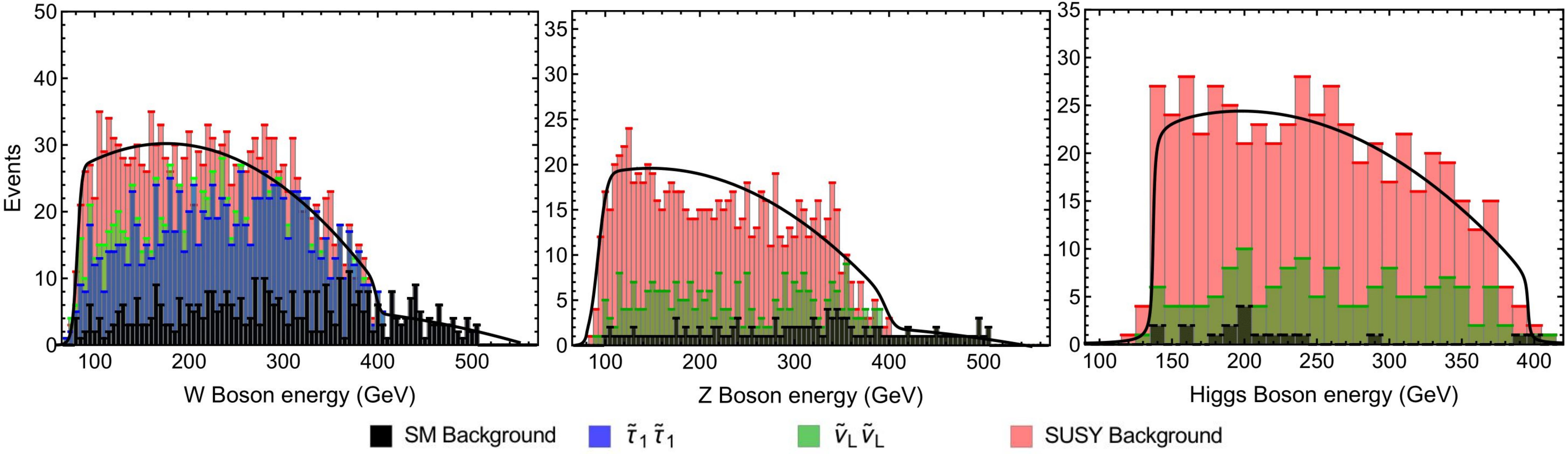}
\caption{Energy spectra of reconstructed $W$ (left), $Z$ (middle) and $h$ bosons (right) for Scenario DEG. SM background is shown in black, while $\tilde\nu_L$ and $\tilde\tau_1$ contributions are shown in green and blue, respectively. Cascade contributions from $\tilde e_{L,R}$, $\tilde\mu_{L,R}$ and $\tilde\tau_2$ are shown in red. The solid line is the result of the fit.}
\label{fig:energyC}
\end{figure}

In case of an observation we would be interested in extracting as much information as possible from the newly observed particles. We aim to evaluate the mass reconstruction via endpoints \cite{Suehara:2009bj} for our three scenarios. For this, we take $m_{\tilde{L}}=300$ GeV, $m_{\tilde{\nu}_R}=100$ GeV as benchmark. In our mass reconstruction we neglect possible effects due to ISR and Beamstrahlung, even though they are included in our simulation, and assume $E_{beam} =500 $ GeV for a luminosity of 500 fb$^{-1}$. We apply the same cuts as before with a stricter condition as  we now require the final state SM bosons to be equal (WW, ZZ, hh). Events fall into three datasets ($W$-like, $Z$-like and $h$-like), based on the type of final state (light jets or b-jets) and the reconstructed invariant masses ($m_W,\,m_Z,\,m_h$). Our channels of interest contribute to this datasets in the following way:
\begin{itemize}
    \item $\tilde{\tau}_1\tilde{\tau}_1\xrightarrow{}$ W-like $\xrightarrow{}$ 4 jets (excluding b-jets)
    \item $\tilde{\nu}_L\tilde{\nu}_L\xrightarrow{}$ Z/h-like $\xrightarrow{}$ 4 jets or 4 bjets
    \item $\tilde{\ell}\tilde{\ell}\xrightarrow{}$ SUSY cascade $\xrightarrow{}$ Z/h like
    
\end{itemize}
We fit the energy spectra of the reconstructed SM bosons for the signal and SM background using a Voigt function convoluted with a second order polinomial. This allows us to obtain the upper and lower endpoints.
The results of the fit are shown on Figure \ref{fig:energyC} for Scenario DEG. Here we see that we can successfully fit the energy spectra for our three SM bosons.

From the 2-body decay kinematics we can obtain the masses as a function of the endpoints. In Table \ref{table:results} we can see the final results for the reconstructed SUSY masses. These are in agreement with the expected value, all within 2$\%$. For Scenario SE, since we lack an on-shell contribution to the W-like dataset it is not possible to reconstruct the mass of the lightest slepton $(m_{\tilde{\ell}_{1}})$. For the other scenarios, $\tilde{\tau}_{1}$ is the lightest slepton, so all three datasets have an on-shell contribution.

\begin{table}[h]
\centering
\setlength{\tabcolsep}{1.5em}
{\begin{tabular}{| c | c | c|  c||c|} 
\hline
Scenario & SE & ST & DEG & Theory \\
\hline
$m_{\tilde\ell_{1}}$(GeV) & - & 296.91 $\pm$ 10.69& 290.51 $\pm$ 10.01&294.47\\
\hline
$m_{\tilde{\nu}_L}$ (GeV) & 293.63 $\pm$ 3.12 & 293.32 $\pm$ 3.61& 293.41 $\pm$ 2.15&293.37\\
\hline
$m_{\tilde{\nu}_R}$ (GeV) & 100.52 $\pm$ 1.65 & 101.14 $\pm$ 1.36& 100.05 $\pm$ 0.67&101.98\\
\hline
\end{tabular}}
\caption{Reconstructed masses in our three scenarios. For $m_{\tilde\ell_1}$, the last column shows the prediction for the lightest stau mass. }
\label{table:results}
\end{table}
\section{Conclusions}
\label{sec:conclusions}

We investigated scenarios of a supersymmetric seesaw model, which can be challenging for the LHC since the final states involve SM bosons. With current bounds as low as 200 GeV, we extended our research to the proposed ILC.

We have shown that a discovery is possible for slepton masses up to 400 GeV with luminosities of 100 fb$^{-1}$, and up 450 GeV for a luminosity of 1 ab$^{-1}$. For higher masses the parameter space suppression becomes too strong. This hold as long as $m_{\tilde l} - m_B - m_{\tilde \nu_R} \gsim 60$~GeV.

Finally, we investigated the prospect of measuring the slepton masses in these scenarios, given that we remain close to our benchmark of $m_{\tilde L}, m_{\tilde E}\sim$ 300 GeV, $m_{\tilde \nu_R}\sim 100$.
 We adapted an endpoint method originally used for the mass measurements of electroweakinos, in scenarios in which these decay dominantly into SM-bosons. For sleptons that decay predominantly on shell, we find that we can reconstruct the masses with a precision of a few percent.

\section*{Acknowledgements}
I would like the thank the co-authors of this work, Nhell Cerna-Velazco, Joel Jones-Perez and Werner Porod, it was great working with you. I would also like to thank the organizers of LCWS2021 for the opportunity to give this talk.

\bibliographystyle{utphys}
\bibliography{SleptonSnuILC}

\end{document}